\documentstyle[12pt]{article}

\renewcommand{\thefootnote}{\fnsymbol{footnote}}
\newcommand{\r}[1]{(\ref{#1})}
\newcommand{\R}{{\sf R\hspace*{-0.9ex}\rule{0.15ex}%
{1.5ex}\hspace*{0.9ex}}}

\begin{document}
\thispagestyle{empty}
\newlength{\defaultparindent}
\setlength{\defaultparindent}{\parindent}

\begin{center}
{\large{\bf TOWARDS A COULOMB GAS OF INSTANTONS OF THE $S0(4) \times 
U(1)$ HIGGS MODEL ON $\R_4$}} 
\vspace{0.5cm}\\

{\large K. Arthur}\\
{\it School of Mathematical Sciences,}\
{\it Dublin City University}\\
{\it Glasnevin, Dublin 9, Ireland.}\vspace{0.5cm}\\

{\large G.M. O'Brien}\\
{\it School of Theoretical Physics,}\
{\it Dublin Institute for Advanced Studies,}\\
{\it 10 Burlington Road,}\
{\it Dublin 4, Ireland.}\vspace{0.5cm}

{\large D.H. Tchrakian\footnote{Supported in part by CEC 
under grant HCM--ERBCHRXCT930362},}\\
{\it Department of Mathematical Physics,}\
{\it St Patrick's College Maynooth,}\
{\it Maynooth, Ireland}\\
{\it School of Theoretical Physics,}\
{\it Dublin Institute for Advanced Studies,}\\
{\it 10 Burlington Road,}\
{\it Dublin 4, Ireland.}\vspace{0.5cm}

\end{center}

\bigskip
\bigskip
\bigskip
\bigskip
\begin{abstract}The $SO(4)\times U(1)$ Higgs model on $\R_4$ is extended 
by a $F^3$ term so that the action receives a nonvanishing contribution 
from the interactions of 2-instantons and 3-instantons, and can be 
expressed as the inverse of the Laplacian on $\R_4$ in terms of the mutual 
distances of the instantons. The one-instanton solutions of both the basic 
and the extended models have been studied in detail 
numerically.\end{abstract}

\vfill
\setcounter{page}0
\renewcommand{\thefootnote}{\arabic{footnote}}
\setcounter{footnote}0
\newpage

\newcommand{\ra}{\rightarrow}

\newcommand{\dd}{\mbox{d}}
\newcommand{\ee}{\end{equation}}
\newcommand{\be}{\begin{equation}}
\newcommand{\ii}{\mbox{ii}}
\newcommand{\pa}{\partial}
\newcommand{\vep}{\varepsilon}
\newcommand{\bfR}{{\bf R}}
\newcommand{\lm}{\lambda}

\pagestyle{plain}

\section{\bf Introduction}
\setcounter{equation}{0}

\parskip0.5truecm

Instantons are expected to play a central role in the semiclassical anaysis 
of non-perturbative phenomena in non-linear field theories. Both the 
Electroweak and the Strong interactions are described by the dynamics of 
the Yang-Mills(YM) model and hence the $SU(2)$ instantons\cite{BPST} of 
the relevant YM models on $\R_4$ in each case are expected to be 
employed. Unfortunately, this programme turned out to be less 
successful than it was hoped. One of the main obstacles encountered was 
the infrared problem arising from large instanton effects in the tunneling 
between topologically distinct vacuua. This problem arises directly as a 
consequence of the scale invariance of the YM model on $\R_4$, which 
results in the dependence of the instanton solution on an {\it arbitrary} 
scale parameter. The introduction of an {\it absolute} scale would overcome 
this problem, and this is the original motivation of the present work.

The most natural way of introducing an {\it absolute} scale is by the 
introduction of Higgs fields, so that the dimensional parameter 
representing the non-zero vacuum expectation value in the Higgs self-
interaction potential can set the {\it absolute} scale of the would be theory. 
A related but additional feature of the presence of a Higgs field is the {\it 
exponential localisation} of the instanton solution, which results in the 
possibility of constructing a multi-instanton field configuaration in which 
the individual instantons overlap only asymptotically. This is what would 
enable the construction of a {\it dilute gas} of instantons. Furthermore 
with the appropriate asymptotic behaviour of these instantons, it may 
become possible to construct a dilute gas of instantons which interact non-
trivially to yield a Coulomb gas in the appropriate dimensions. By a 
Coulomb gas we understand a gas where the contribution to the {\it action} 
coming from the interactions is given by the inverse of the Laplacian in 
terms of the mutual distances of the constituents of the gas. It is our aim in 
the present work to propose a gauged Higgs model on $\R_4$ which 
acheives the objectives just stated.

The construction of such a Coulomb gas for a Higgs model on $\R_3$ was 
performed long ago by Polyakov\cite{P}. The Higgs model there\cite{P} 
consisted of the $SU(2)$ YM field interacting with an $su(2)$ valued Higgs 
field and the usual symmetry breaking Higgs potential. Firstly, the well 
known monopole\cite{'tHP} solution of this model was taken to be the {\it 
exponentially} localised instantons in $2+1$ dimensions. Secondly, a 
Coulomb gas of these instantons was constructed using the asymptotic fields 
of the monopole\cite{'tHP}. Our plan in the present work is to propose a 
gauged Higgs model which supports instanton solutions with the requisite 
asymptotic properties capable of describing a Coulomb gas on $\R_4$. We 
shall restrict ourselves to this first task here, and defer the second 
technically complicated task of constructing the resulting dilute gas action 
to a future work. Before proceding we make a remark aimed at putting the 
task at hand in perspective: Polyakov's construction\cite{P} of the Coulomb 
gas of instantons on $\R_3$ is the $3$ dimensional analogue of the $2$ 
dimensional Coulomb gas of instantons contructed previously by 
Berezinsky and by Kosterlitz and Thouless\cite{B-KT} employing the $O(2)$ 
model on $\R_2$, while the present work proposes the corresponding 
instanton field configuration on $\R_4$.

Our sole task in this paper is to find the instanton solution of a particular 
Higgs model satisfying the above stated criteria. We shall not attempt to 
compute the action of the corresponding Coulomb gas. Our plan is based on 
the $3$ dimensional example\cite{P}, namely to construct an appropriate 
gauged Higgs model in contrast with the non-gauged model employed in 
the $2$ dimensional example\cite{B-KT}. Section {\bf 2} will be concerned 
with the instanton solution of the basic $SO(4) \times U(1)$ Higgs model 
proposed in Ref.\cite{O'BT}. We will make a detailed numerical study of this 
solution and find out that inspite of the instanton in question being 
exponentially localised, the asymptotic properties do not permit the 
construction of a dilute gas whose inter-instanton interactions support a 
Coulomb gas. Then in Section{\bf 
3} we shall propose two extended versions of the basic model and will verify 
the existence of unit topological charge instantons by numerical 
techniques. These extended models are designed 
specially to support a non-trivial dilute Coulomb gas on $\R_4$. Finally in 
Section {\bf 4} we will summarise our results and give a brief discussion of 
the outlook for the extended models.

\section{\bf The basic model}

First we briefly describe the $SO(4)\times U(1)$ model introduced in 
Ref.\cite{O'BT} and then procede to subsections {\bf 2.1}, {\bf 2.2} and {\bf 
2.3} where we give the asymptotic solutions, the numerical solutions and, 
an analysis of the inter-instanton interactions, respectively.

The basic model on $\R_4$ is described by the antihermitian $SO(4)\times 
U(1)$ gauge connection $\hat A_{\mu} =A_{\mu} +i\gamma_5 a_{\mu}$ 
and the antihermitian Higgs multiplet $\Phi =\gamma_5 \gamma_a 
\phi_a$, $A_{\mu}$ being the $so(4)$ connection, $a_{\mu}$ the $U(1)$ 
connection and $\phi_a$ a 4-vector real Higgs field. The model, which is 
derived from the $8$-dimensional member of the scale invariant Yang-
Mills hierarchy\cite{T1} on $\R_4 \times S^4$ by dimensional 
descent\cite{T2}, has the following Lagrange density

\[
{\cal L}_4 =\mbox {Tr} [ \hat F_{\mu \nu \rho \sigma}^2 +4\lambda_1 \{ 
\hat 
F_{[\mu \nu},\hat D_{\rho ]} \Phi \} ^2-18\lambda_2 (\{ (\eta^2 
+\Phi^2),\hat F_{\mu \nu} \} -[\hat D_{[\mu} \Phi ,\hat D_{\nu ]} \Phi])^2 
\]
\begin{equation}
\label{1}
-54\lambda_3 \{ (\eta^2 +\Phi^2),\hat D_{\mu}\Phi  \} ^2 +54\lambda_4 
(\eta^2 +\Phi^2)^4 ]
\end{equation}

\noindent
In \r{1} and everywhere below, the brackets $[,]$ and $\{ ,\}$ denote 
commutators and anticommutators respectively, and square brackets 
around indices signify total antisymmetrisation. Also the curvature $\hat 
F$ and the covariant derivative $\hat D$ pertain to the $SO(4)\times U(1)$ 
connection $\hat A = A+i\gamma_5 a$, while below we shall use $F$ and $D$ 
pertaining to the $SO(4)$ connection $A$. The 4-form curvature field 
$F_{\mu \nu \rho \sigma}$, in this notaion, is defined as $F_{\mu \nu \rho 
\sigma} =\{ F_{\mu [\nu},F_{\rho \sigma]} \}$. The last term in \r{1} 
multiplying $\lambda_4$ is the Higgs symmetry breaking potential. It is 
clear that any one of the four dimensionless coupling constants 
$\lambda_a,\: \: a=1,2,3,4$, which we shall denote with a vector notation 
$\vec \lambda$, may be set equal to $1$ by rescaling the $\Phi$ 
field, but we do not do that here.

A physically very important property of the system \r{1} is, that at high 
momenta or small distances when the magnitude of the Higgs field is 
negligible with respect to the dimensional constant $\eta$, the system is 
dominated by the term with the highest power of $\eta$. This happens to be
\begin{equation}
\label{1a}
{\cal L}_{high} \sim \mbox {Tr} [ -\frac{\lambda_2}{3} \hat F_{\mu \nu}^2 
-\lambda_3 (\hat D_{\mu} \Phi)^2 +\frac{\lambda_4}{4} (\eta^2 
+\Phi^2)^2],
\end{equation}
namely the usual Yang-Mills-Maxwell-Higgs system which appears as the 
coefficient of the $\eta^4$ term in \r{1}, which is its dominant 
contribution at high energy leading to a perturbative action as expected. 
Thus the model \r{1} can be regarded as the corresponding effective action 
density. 

There are two other physically relevant properties of the system \r{1}. The 
first is that the gauge group $SO(4)$ leads to a chirally symmetric 
dynamics, hence does not fit in with Electroweak theory. The other is that 
at low momentum or large distances, the gauge field dynamics is dominated 
by the 4-form curvature $\hat F_{\mu \nu \rho \sigma}^2$ whose 
propagation properties again do not fit in with Electroweak dynamics. We 
must therefore conclude that the model \r{1} is not suited to be a prototype 
for an Electroweak theory, but it is consistent with the features Strong 
interaction dynamics both at high and low momenta. In this respect, the 
$U(1)$ connection $a_{\mu}$ in the gauge field multiplet $\hat A 
=(A_{\mu} +i\gamma_5 a_{\mu})$ is irrelevant and can consistently be 
suppressed. We have left $a_{\mu}$ in the definition of ${\cal L}_4$ in 
\r{1}, because that is the most general\cite{ST} model descending from the 
$4p$ dimensional member of the Yang-Mills hierarchy that can support 
instantons, and also because when we restrict to the spherically symmetric 
configuration below the imposition of this symmetry will eliminate this 
$U(1)$ field.

In view of the above reasoning, we will study the properties of this model 
in the context of Strong interaction dynamics at low momentum, and in 
particular will attempt to construct a dilute gas of instantons that satifies 
the properties of a Coulomb gas in $\R_4$, analogously to Polyakov's 
work\cite{P} in $\R_3$. As the title states however, this programme will 
not be completed but we will restrict henceforth to the study of the 
classical instanton solutions of this model and its extensions.

When the values of $\lambda_a$ respect the following restrictions: 
$\lambda_a > 1$ for each $a$, then the action density \r{1} is bounded from 
below by the topological charge density $\varrho = \partial_{\mu} 
\Omega_{\mu}$, expressed as the divergence of the Chern-Simons form 
$\Omega_{\mu}$ given by\cite{O'BT}
\[
\Omega_{\mu} =-\frac{1}{8\pi^2} \varepsilon_{\mu \nu \rho \sigma} 
\mbox {Tr}\gamma_5 [\eta^4 \hat A_{\nu} (\hat F_{\rho \sigma} 
-\frac{2}{3} \hat 
A_{\rho} \hat A_{\sigma})
\]
\begin{equation}
\label{2}
-\frac{1}{6} \eta^2 \Phi \{ \hat F_{[\rho \sigma},\hat D_{\nu ]} \Phi \}  - 
\frac{1}{6} \Phi (\{ (\eta^2 +\Phi^2),\hat F_{\rho \sigma} \} +\hat 
D_{[\rho} \Phi \: \: \hat D_{\sigma]} \Phi)\hat D_{\nu}\Phi ].
\end{equation}
The surface integral of this density is the topological charge $N$, and the 
spherically symmetric solutions with which we will be concerned in this 
paper yield $N=1$ for this integral.

In the limit where $\lambda_a =1$ for all $a$, or $\vec \lambda =(1,1,1,1), 
$this bound is saturated and the 
action is equal to the topological charge $N$. If however any of the 
$\lambda_a$ take positive values smaller than $1$, then the action is not 
bounded any more by $N$ but by $\lambda_{min} N$, where 
$\lambda_{min}$ is the smallest of those $\lambda_a$ which are 
smaller than $1$. This can be shown using similar arguments to those used 
in Ref.\cite{JbR} in the case of the Abelian Higgs model. When $\lambda_a 
<0$ for any $a$, we lose the topological 
lower bound. 

In the case where the topological lower bound is saturated we have the 
following self-duality, or Bogomol'nyi, equations: 
\begin{equation}
\label{3}
\frac{1}{36} \varepsilon_{\mu \nu \rho \sigma} \gamma_5 \hat F_{\mu 
\nu \rho \sigma} -(\eta^2 +\Phi^2)^2 =0
\end{equation}
\begin{equation}
\label{4}
\frac{1}{9} \varepsilon_{\mu \nu \rho \sigma} \gamma_5 \{ \hat F_{[\rho 
\sigma} ,\hat D_{\mu]} \Phi \} -\{ (\eta^2 +\Phi^2),\hat D_{\nu} \Phi \} =0
\end{equation}
\begin{equation}
\label{5}
\frac{1}{2} \varepsilon_{\mu \nu \rho \sigma} \gamma_5 (\{ (\eta^2 
+\Phi^2),\hat F_{\rho \sigma} \} +\hat D_{[\rho} \Phi \: \: \hat 
D_{\sigma]} \Phi) - (\{ (\eta^2 +\Phi^2),\hat F_{\mu \nu} \} +\hat 
D_{[\mu} \Phi \: \: \hat D_{\nu]} \Phi) =0.
\end{equation}
The system \r{1} shares an important property with the Skyrme\cite{S} 
model on $\R_3$, namely that all but the second power of the derivative of 
any field component are excluded. This has been a criterion in the 
construction of \r{1}. Like the Skyrme model also, the self-duality 
equations saturating the topological bound are 
overdetermined\cite{O'BT}\cite{TC}. Accordingly the only solutions of 
the equations \r{3}-\r{5} are the trivial constant solutions we shall see 
below, and any non-trivial solutions we find will satisfy the second order 
Euler-Lagrange equations and not \r{3}-\r{5} even for $\lambda_i =1$ for 
all $i$.

The most important property we shall require of the instanton solution is 
the large $r$ limit of the magnitude of the Higgs field $|\Phi|=\sqrt{-
\mbox{Tr} \Phi^2}$
\begin{equation}
\label{6}
\lim_{r \rightarrow \infty} |\Phi | =\eta
\end{equation}
in terms of $\eta$, the vacuum expectation value of the Higgs field. 
Requiring this asymptotic condition, we shall find out that the 
corresponding behaviours of the $SO(4)$ gauge connection $A_{\mu}$ and 
the full Higgs multiplet are
\begin{equation}
\label{7}
\lim_{r \rightarrow \infty} A_{\mu} =\frac{1}{2} g^{-1} \partial_{\mu} \: 
g\: ,\qquad \qquad \lim_{r \rightarrow \infty} \Phi = \eta g,
\end{equation}
in terms of the $SO(4)$ group element $\gamma_5 \gamma_{\mu} \hat 
x_{\mu}$, with $\hat x =
\frac{x}{r} $. We now adapt the arguments in Ref.\cite{JR} 
pertaining to the $SU(2)$ YM instanton to the solutions satisfying \r{7}. In 
the temporal gauge $A_0 =0$ each component of $A_{\mu}$ and $\Phi$ 
become time independent, thus enabling the identification of our solutions 
as instantons.

The solutions we shall seek below are restricted to the spherically 
symmetric restriction of the Euler-Lagrange equations, namely the unit 
topological charge instantons. This restriction is quite adequate since 
instanton configurations of all chrges can be attained by considering the 
collection of arbitrarily many $N=1$ intantons, since as we shall see below 
these overlap only asymptotically. The reason is the presence of the Higgs 
field which introduces the absolute scale with respect to which the 
instantons will be localised. Indeed, we shall be able to acheive an {\it 
exponential} decay yielding this localisation, similar to the instantons of 
the $\phi^4$ model\cite{P2} on $\R_2$ and 'compact 
electrodynamics'\cite{P}\cite{P2} on $\R_3$.

\subsection{Asymptotic solutions}

Since we are unable to solve analytically the Euler-Lagrange equations of 
the one-dimensional subsystem of \r{1} arising from the imposition of 
spherical symmetry, we will give explicit solutions in the $r>>1$ and $r<<1$ 
asymptotic regions only.

Under the imposition of spherical symmetry on $\R_4$, the Abelian field 
$f_{\mu \nu} =\partial_{[\mu} a_{\nu]}$ defining the system \r{1} 
vanishes and we are left with only the $SO(4)$ field $\hat A_{\mu} = 
A_{\mu}$, for which we employ the following spherically symmetric 
Ansatz

\begin{equation}
\label{8}
A_{\mu}=\frac{1+f(r)}{r} \gamma_{\mu \nu} \hat x_{\nu} ,\qquad \qquad 
\Phi =h(r)\: \gamma_5 \gamma_{\mu} \hat x_{\mu} ,
\end{equation}

\noindent
where $f(r)$ is a dimensionless function and $h(r)$ has the same 
dimensions as $\eta$. $\gamma_{\mu \nu} =-\frac{1}{4} [\gamma_{\mu} 
,\gamma_{\nu}]$ is the (Dirac) spinor representation of $SO(4)$.

\noindent
Substituting \r{8} into \r{1}, performing the angular integrations and 
with a convenient rescaling, we have the following one-dimensional 
subsystem

\[
L_4 =\frac{1}{r^3} (1-f^2)^2 f_r^2 +\frac{4\lambda_3}{3} r^3 (\eta^2 -h^2)^2 
h_r^2
\]
\[
+\frac{\lambda_1}{r} \left( [(1-f^2)h]_r\right)^2 +\lambda_2 r \left( 
[(\eta^2 -h^2)f]_r \right)^2
\]
\[
+\frac{3\lambda_1}{r^3} \left( fh(1-f^2)\right)^2 +4\lambda_3 r\left( 
fh(\eta^2 -h^2)\right)^2
\]
\begin{equation}
\label{9}
+\frac{\lambda_2}{r} \left( (1-f^2)(\eta^2 -h^2)+2f^2 h^2 \right)^2 
+\frac{\lambda_4}{4} r^3 (\eta^2 -h^2)^4,
\end{equation}
having suppressed the overall constant factor coming from the angular 
volume. The subscript $r$ in \r{9} 
denotes the differentiation $f_r =\frac{df}{dr}$.

\noindent
Using the notation $\delta_f L =\frac{\partial L}{\partial f} -\frac{d}{dr} 
(\frac{\partial L}{\partial f_r})$, we express the two Euler-Lagrange 
equations of \r{9} . The equation $\delta_f L =0$ arising from the arbitrary 
variations $\delta f$ is
\[
\delta_f L_4 =\frac{2}{r^3} (1-f^2)\left( 2f f_r^2 +\frac{3}{r} (1-f^2)f_r -(1-
f^2)f_{rr} \right)
\]
\[
+\frac{4\lambda_1}{r^2} fh\left( [(f^2 -1)h]_r -r[2f_r^2 h+2f f_{rr} h 
+4ff_r h_r +(f^2 -1)h_{rr}] \right)
\]
\[
-2\lambda_2 (h^2 -\eta^2)\left( [(h^2 -\eta^2)f]_r +r[2h_r^2 f +2hh_{rr} f 
+4hh_r f_r +(h^2 -\eta^2)f_{rr}]\right)
\]
\[
+\frac{6\lambda_1}{r^3} fh^2 (f^2 -1)(3f^2 -1) +8\lambda_3 rfh^2 (h^2 
-\eta^2)^2
\]
\begin{equation}
\label{10}
+\frac{4\lambda_2}{r} f(3h^2 -\eta^2)\left( (h^2 -\eta^2)(f^2 -1) +2f^2 h^2 
\right) ,
\end{equation}
and using the similar notation $\delta_h L =\frac{\partial L}{\partial h} 
-\frac{d}{dr} (\frac{\partial L}{\partial h_r})$, the equation $\delta_h L 
=0$ arising from $\delta h$ is given by
\[
\delta_h L_4 =-\frac{8\lambda_3}{3} r^2 (h^2 -\eta^2)\left( 2rhh_r^2 +(h^2 
-\eta^2)(rh_{rr} +3h_r)\right)
\]
\[
+\frac{\lambda_1}{r^2} (f^2 -1)\left( [(f^2 -1)h]_r -r[2f_r^2 h+2ff_{rr} 
h+4ff_r h_r (f^2 -1)h_{rr}]\right)
\]
\[
-4\lambda_2 hf\left( [(h^2 -\eta^2)f]_r +r[2h_r^2 f+2hh_{rr} f+4hh_r f_r 
+(h^2 -\eta^2)f_{rr}]\right)
\]
\[
+\frac{6\lambda_1}{r^3} (f^2 -1)^2f^2 h+24\lambda_3 rf^2 h(h^2 
-\eta^2)(3h^2 -\eta^2)
\]
\begin{equation}
\label{11}
+\frac{4\lambda_2}{r} h(3f^2 -1)\left( (h^2 -\eta^2)(f^2 -1) +2h^2 f^2 
\right) +2\lambda_4 r^3 (h^2 -\eta^2)^3 h
\end{equation}

Since the system \r{1} is a non Abelian Higgs model like the models 
employed in Refs.\cite{'tHP}, we expect the solution to have asymptotic 
behaviour similar to that of the monopole\cite{'tHP}. Accordingly we seek 
solutions with the following properties
\begin{equation}
\label{12}
\lim_{r \rightarrow 0} f(r)=-1 \qquad \qquad  \lim_{r \rightarrow \infty} 
f(r)=0
\end{equation}
\begin{equation}
\label{13}
\lim_{r \rightarrow 0} h(r)= 0  \qquad \qquad  \lim_{r \rightarrow 
\infty} h(r)=\eta .
\end{equation}
The first members of both \r{12} and \r{13} guarantee that the solution is 
regular at $r=0$, while the second members satisfy criteria of finite action. 
Moreover, the second members of these equations result in the asymptotic 
behaviour anticipated in \r{7}, which is essential for the interpretation of 
the finite action topologically stable solution as the instanton of the 
dynamical model.

Expanding around their asymptotic values $f(r)=0+F(r)$ and $h(r)=\eta 
-H(r)$ in the region $r>>1$ in terms of the small functions $F(r)$ and 
$H(r)$, and retaining only linear terms in these functions, \r{10} and 
\r{11} reduce respectively to
\begin{equation}
\label{14}
\rho^2 F_{\rho \rho} -3\rho F_{\rho} -3\rho^2 F=0
\end{equation}
\begin{equation}
\label{15}
\sigma^2 H_{\sigma \sigma} -\sigma H_{\sigma} -8\sigma^2 H=0,
\end{equation}
in which we have used the following two dimensionless rescalings of $r$: 
$\rho =\sqrt{\lambda_1} \eta r$, $\sigma 
=\sqrt{\frac{\lambda_2}{2\lambda_1}} \eta r$.

\noindent
The solutions of \r{14} and \r{15} respectively are expessed in terms of 
modified Bessel functions and lead to the following {\it exponentially} 
decaying solutions
\begin{equation}
\label{16}
f(r) =\lambda_1 \eta^2 r^2 K_2 (\sqrt{3\lambda_1} \eta r)
\end{equation}
\begin{equation}
\label{17}
h(r)=\eta \left( 1 -\sqrt{\frac{\lambda_2}{2\lambda_1}} \eta r K_1 
(2\sqrt{\frac{\lambda_2}{\lambda_1}} \eta r) \right) ,
\end{equation}
in the $r>>1$ region.

In the $r<<1$ region, we tried a solution in powers of the rescaled radial 
variable $\rho= \eta r$ and found

\begin{equation}
\label{18}
f(\rho ) =A\rho^2 +o(\rho^4)
\end{equation}
\begin{equation}
\label{19}
h(\rho ) =B\rho +o(\rho^3)
\end{equation}

\noindent
where the dimensionless constant $A$ and the constant $B$ with the 
dimensions of $\eta$ are arbitrary and will be determined by the 
numerical integrations immediately below. Note that the behaviours \r{18} 
and \r{19} lead to fields $(A_{\mu},\Phi)$ \r{8} that are regular and 
differentiable at $r=0$.

\subsection{Numerical solutions}

We have integrated the equations $\delta_f L_4 =0$ and  $\delta_h L_4 =0$ 
given respectively by \r{10} and \r{11} numerically starting with 
the functions \r{18} and \r{19} near $r=0$ and choosing the numerical 
values of the constants $A$ and $B$ so that $f(r)$ and $h(r)$ 
reach their asymptotic values given by the second members of \r{12} and 
\r{13}.  We have performed this integration for the following values of the 
dimensionless coupling constants $\vec \lambda_{(i)} =(\lambda_1 
,\lambda_2 ,\lambda_3 ,\lambda_4 )$, with $\vec \lambda^{(1)} 
=(0.5,0.5,0.5,0.5)$, $\vec \lambda^{(2)} =(0.8,0.8,0.8,0.8)$, $\vec 
\lambda^{(3)} =(1,1,1,1)$, and $\vec \lambda^{(4)} =(1.2,1.2,1.2,1.2)$. The 
numerical values of the pair of constants $\{ A, B\}$ are fixed by each of 
these the numerical integrations, and are listed respectively in the first 
and second columns of Table 1. The profiles of the functions $f(r)$ are 
given in Figure 1, and those of $h(r)$ are given in Figure 2. We do not 
exhibit the profiles of the action densities corresponding to these solutions 
since they are all ball shaped as expected of spherically symmetric lumps. 
The values of the action integrals $S(\vec 
\lambda^{(i)})$, $i=1,2,3,4$ for each of these solutions are listed in the third 
column of Table 1. We defer a detailed discussion of these quantitative 
results to Section {\bf 4}.

The above integrations demonstrate the existence of the spherically 
symmetric solutions of the {\it basic} model\cite{O'BT}, but we find it 
interesting to pursue our numerical studies somewhat further because of 
an unusual feature of the model at hand. This concerns the 
overdetermined\cite{TC} nature of the Bogomol'nyi equations \r{3}-\r{5}, a 
feature shared with with the Skyrme model\cite{S} on $\R_3$. While in the 
usual Skyrme model the overdetermined self-duality equations are 
parametrised only by one dimensional constant which sets the absolute 
scale, here in addition to the corresponding dimensional constant $\eta$ we 
have the three independent components of the four dimensionless 
coupling constants $\vec \lambda$. It is therefore interesting to 
investigate numerically the quantitative departure from the 'minimal' 
configuration $\vec \lambda =(1,1,1,1)$.

We know from Refs.\cite{O'BT}\cite{TC} that \r{3}-\r{5} are only satisfied 
by the constant asymptotic values of $f(r)$ and $h(r)$, given by \r{12} and 
\r{13}, and therefore that the stress tensor ${\cal T}_{\mu \nu}$ 
corresponding to these solutions does not vanish. It is interesting to see 
how the component ${\cal T}_{44}$ of the stress tensor behaves in detail, 
quantitatively. Now the component $T_{44}$ of the stress tensor 
corresponing to the spherically symmetric sub-system equivalent to the 
one-dimensional Lagrangian \r{9} is

\[
T_{44} =\left( \frac{1}{r^3} (1-f^2)^2 f_r^2 -\frac{\lambda_4}{4} r^3 (\eta^2 
-h^2)^4\right) +\left( \frac{4\lambda_3}{3} r^3 (\eta^2 -h^2)^2 h_r^2 
-\frac{3\lambda_1}{r^3} f^2 h^2 (1-f^2)^2 \right)
\]
\[
+\left( \frac{\lambda_1}{r} [(1-f^2)h]_r^2 -4\lambda_1 r f^2 h^2 (\eta^2 
-h^2) \right)
\]
\begin{equation}
\label{20}
+\left( \lambda_2 r [(\eta^2 -h^2)f]_r^2 -\frac{\lambda_2}{r} [(1-
f^2)(\eta^2 -h^2)+2f^2 h^2]^2 \right).
\end{equation}

\noindent
Note that each one of the four large brackets in\r{20} vanishes 
separately in the (overdetermined) self-dual limit $\vec \lambda 
=(1,1,1,1)$. In that case, the vanishing of each large bracket states the 
corresponding Bogomol'nyi equation, namely

\begin{equation}
\label{21}
\frac{1}{r^3} (1-f^2)f_r +\frac{1}{2} (\eta^2 -h^2) =0
\end{equation}
\begin{equation}
\label{22}
2 (\eta^2 -h^2) h_r -\frac{3}{r^3} (1-f^2)fh =0
\end{equation}
\begin{equation}
\label{23}
\frac{1}{r^2} [(1-f^2)h]_r +2(\eta^2 -h^2)fh =0
\end{equation}
\begin{equation}
\label{24}
\frac{1}{r} [(\eta^2 -h^2)f]_r -\frac{1}{r^2} [(1-f^2)(\eta^2 -h^2) +2f^2 h^2] 
=0.
\end{equation}
Thus $T_{44}$ \r{20} gives a quantitative measure of the departure of a 
solution from the trivial self-dual field configuration satisfying \r{21}-
\r{24}.We have plotted the profiles of $T_{44}$ for each of the four 
solutions exhibited in Table 1 and Figures 1-3, in Figure 4.

Equations \r{21}-\r{24} are the spherically symmetric restrictions of the 
Bogomol'nyi equations \r{3}-\r{5}. Specifically, \r{21} pertains to \r{3}, 
\r{22} and \r{23} to \r{4} and \r{24} to \r{5}. It can be observerved 
immediately that when each of \r{21}-\r{24} is squared, the sum of the 
square terms yields the spherically symmetric restriction of \r{1}, namely 
the one-dimensional system \r{9}. The corresponding sum $\sigma$ of the 
cross terms is then the spherically symmetric restriction of the topological 
charge density $\varrho = \partial_{\mu} \Omega_{\mu}$ \r{2}, divided by 
$\frac{\rho^3}{8\pi^2 \eta^3}$. Since $\varrho$ is a total divergence, the 
corresponding one dimensional topological charge density which is this 
quantity $\sigma$, turns out to be a total derivative given by

\begin{equation}
\label{25}
\sigma =-\frac{1}{2}\frac{d}{d\rho} \left( 
3f(1-\frac{1}{3} f^2) -6(1-f^2)fg^2 +(1-3f^2)fg^4 \right) ,
\end{equation}

\noindent
which is expressed in terms of the rescaled radius $\rho =\eta r$ and the 
rescaled dimensionless function $g(\rho) =\eta^{-1} h(\rho)$. The integral 
$\int \sigma d\rho$, the topological charge of the one-dimensional 
subsystem \r{9}, is immediately evaluated using the limits \r{12} and \r{13} 
to yield $N=1$.

Having acheived a detailed quantitative understanding of our spherically 
symmetric solutions, we procede to consider the question of inter-instanton 
interactions.

\subsection{Inter-instanton interaction}

As we saw above, even for the coupling constant vector $\vec \lambda 
=(1,1,1,1)$, the instanton solution is {\it not} self-dual and the stress tensor 
does {\it not} vanish. It is therefore expected that the force between such 
instantons is non-vanishing, and hence that a dilute gas of such instantons 
can be constructed.

In this paper, we shall adapt Polyakov's\cite{P} construction of a dilute gas 
of instantons on $\R_3$ to our problem on $\R_4$. As in Ref.\cite{P}, we 
will set out to construct a Coulomb gas of intantons. Given that our solution 
is expressed by the Ansatz \r{8} and obeys boundary conditions \r{12} and 
\r{14}, we opt to work in the Dirac-string gauge introduced in 
Refs.\cite{O'BT}\cite{MT}. In this gauge, the Higgs field in the $r>>1$ region 
$\Phi(\infty) = \eta \gamma_5 \gamma_{\mu} \hat x_{\mu}$ is gauged to 
be a constant valued field $^{\omega} \Phi = \eta \gamma_5 \gamma_4$, by 
rotating it in the direction of the $x_4$ axis under the action of a suitable 
$SO(4)$ gauge rotation $\omega$. As a result the Higgs potential and the 
covariant derivative of the Higgs field vanish, reducing the (asymptotic 
$r>>1$) action density \r{1} to the only remaining term $F_{\mu \nu \rho 
\sigma}^2$.

The gauge field connection and curvature in the Dirac-string 
gauge are calculated from \r{8} with $f(r)=0$ according to \r{12}. Just as 
the $SU(2)$ symmetry of the connection in the Georgi-Glashow model on 
$\R_3$ breaks to $U(1)$ for the asymptotic fields in the Dirac string gauge, 
so the $SO(4)$ symmetry of the connection \r{8} breaks to $SO(3)$ here. 
This $SO(3)$ valued
asymptotic connection and its curvature are given\cite{O'BT}\cite{MT}, 
respectively, by

\begin{equation}
\label{26}
^{\omega}A_i =\frac{1}{(1+\hat x_4)r} \gamma_{ij} \hat x_j \: \: ,\qquad 
\qquad \qquad ^{\omega}A_4 =0,
\end{equation}
\begin{equation}
\label{27}
^{\omega}F_{ij}=-\frac{1}{r^2} \left( \gamma_{ij} +\frac{1}{1+\hat x_4} 
\hat x_{[i} \gamma_{j]k} \hat x_k \right) ,\qquad \qquad 
^{\omega}F_{i4}=\frac{1}{r^2} \gamma_{ij} \hat x_j ,
\end{equation}

\noindent
where the same notations as above are used, and with the index $\mu =i,4, \: 
\: i=1,2,3$. The Dirac-string singularity along the negative $x_4$ axis is 
manifest in \r{26}. Indeed the $SO(d)$ gauge field on $\R_d$ for 
arbitrary $d$ breaks down to an $SO(d-1)$ asymptotic field in the Dirac-
string gauge, given by \r{27} in which the 
subscripts $4$ are replaced by $d$, the gamma matrices are understood to 
be the $d$-dimensional gamma matrices and the singularity in the 
connection is 
on the negative $x_d$-axis. This is explained in detail in Ref.\cite{MT}. Note 
that the apparent lack of rotational invariance in 
\r{27} is just a gauge artifact, and it is easy to check that gauge invariant 
quantities such as $\mbox {Tr}\: F_{\mu \nu}^2 $, $\mbox {Tr} ^{*}F_{\mu 
\nu} \: F_{\mu \nu}$, $\mbox{Tr} F_{\mu \nu}F_{\nu \rho}F_{\rho \mu}$ 
etc., are $SO(4)$ scalars. Only in the $d=3$ case does \r{27} take a 
manifestly $SO(3)$ invariant form, namely
\[
^{\omega}F_{\mu \nu} =\frac{1}{2r^2} \varepsilon_{\mu \nu \lambda} 
\hat x_{\lambda} \: \: \sigma_3 .
\]
The components of the curvature in \r{27} are given in Cartesian 
coordinates and are valid in the region $r>>1$. It is therefore desirable to 
express the components of the curvature on the 3-sphere $S^3$ at infinity 
in polar coordinates like the Wu-Yang monopole\cite{WY} whose only non-
vanishing component on the 2-sphere $S^2$ at infinity is 
$^{\omega}F_{\theta \phi} =\frac{1}{2} \sigma_3 \: \sin \theta$. In terms 
of the two polar coordinates $\psi$ and $\theta$, and the azimuthal 
coordinate $\phi$ on $S^3$ at infinity, the non-vanishing components of 
the curvature $^{\omega}F$ are

\begin{equation}
\label{28}
^{\omega}F_{\psi \theta}=\sin \psi (\gamma_{31} \cos \phi +\gamma_{23} 
\sin \phi),
\end{equation}
\begin{equation}
\label{29}
^{\omega}F_{\theta \phi}=-\sin^2 \psi \sin \theta \left( \gamma_{12} \cos 
\theta +(\gamma_{31} \sin \phi +\gamma_{23} \cos \phi)\sin \theta 
\right),
\end{equation}
\begin{equation}
\label{30}
^{\omega}F_{ij}=-\sin \psi \sin \theta \left( -\gamma_{12} \sin \theta 
+(\gamma_{31} \sin phi +\gamma_{23} \cos \phi) \cos \theta \right),
\end{equation}

Since our instantons are exponentially localised, c.f. \r{16} and \r{17}, we 
shall consider that a gas of non overlapping instantons at positions $\{ x_a 
\}$, with $x_{ab} =|x_a^{\mu} -x_b^{\mu}| >> \eta^{-1}$ can be described by 
the linear superposition

\begin{equation}
\label{31}
F_{\mu \nu}= \sum_{a} q_a \: \: ^{\omega}F_{\mu \nu}(x-x_a), \: \: \: (q_a 
=\pm 
1).
\end{equation}
Following Ref.\cite{P}, we argue that the first contribution to the action 
comes from the sum over $a$ of the action integrals of the one-instanton 
solution. The first, dominant contributions to these integrals comes from 
the 
integration of the action density
\begin{equation}
\label{32}
{\cal L}_4^{(1)} =\sum_{a} \mbox Tr \{ F_{\mu [\nu}(x-x_a),F_{\rho 
\sigma]}(x-x_a) \}^2 ,
\end{equation}
where the integration is restricted to {\it inside} the 4-dimensional spheres 
of radii $R$ around each instanton, such that $\eta^{-
1}<<R<<x_{ab}$. The second contribution comes from the inter-instanton 
interaction terms in the action density calculated from \r{31}, integrated 
over the 4-dimensional volume {\it outside} these spheres with radii $R$. 
The second contribution to the action is the integral of the following 
density
\begin{equation}
\label{33}
{\cal L}_4^{(2)}=\sum_{a,b,c,d} \mbox Tr \{ F_{\mu [\nu}(x-x_a),F_{\rho 
\sigma]}(x-x_b) \}\{ F_{\mu [\nu}(x-x_c),F_{\rho \sigma]}(x-x_d) \},
\end{equation}
where at most two of the four summation indices $a,b,c$ and $d$ must be 
different. Otherwise, when $a=b=c=d$, the sum is over one index only and 
\r{32} and \r{33} 
coincide, and the integral in the region {\it outside} $R$ is negligible 
compared with that in the region {\it inside} $R$.

It turns out that when the asymptotic field strengths \r{27} in the Dirac-
string gauge are subsituted into \r{33} the later vanishes exactly, leading 
to vanishing contribution to the action due to inter-instanton interactions. 
This is because the 4-form curvature $F_{\mu \nu \rho \sigma}$ 
constructed from two distictly situated 2-form curvatures $F_{\mu \nu}(x-
x_a)$ and $F_{\rho \sigma}(x-x_b)$ itself vanishes exactly.
This last statement can be understood more succinctly by noting that the 
asymptotic curvature 
2-form given by \r{28}-\r{30} is defined with respect to the {\it three} 
coordinates $\{ \psi , \theta , \phi \}$ on $S^3$ and to construct a non-
vanishing curvature 4-form $^{\omega}F_{ABCD}$ we need at least {\it 
four} coordinates.

\section{The extended models}

The status of the unit charge spherically symmetric instanton solution of 
the {\it basic} $SO(4) \times U(1)$ model \r{1} is that, while the individual 
lumps are exponentially localised permitting the construction of a dilute 
instanton gas, the asymptotic interactions of these instantons are not 
sufficiently strong to lead to a Coulomb gas, in contrast to the situation in 
the two well known models, namely the $O(2)$ model in {\it two}\cite{B-KT} 
and the Georgi-Glashow model in {\it three}\cite{P} 
dimensions. Since it is our eventual aim to construct a Coulomb gas of 
instantons in {\it four} dimensions, the {\it basic} model \r{1} must be 
modified. The search for such possible modifications is the purpose of this 
Section. The asymptotic solutions of these will be given in Subsection {\bf 
3.1} and their numerical solutions in Subsection {\bf 3.2}.

The modification of \r{1} will take the form of an extension of \r{1}. The 
extended model will consist of \r{1} plus some other gauge invariant terms 
depending on $(A_{\mu},\Phi)$, such that at least one term in it depends 
only on the curvature $F_{\mu \nu}$ and is independent of the Higgs field 
$\Phi$. This is because in the Dirac-string gauge \r{26} all gauge invariant 
quantities depending on $\Phi$ vanish. If the extension consists of a 
positive definite quantity, then it would be expected that the topological 
lower bound given by the surface integral of \r{2} remains valid and 
hence that the new instanton will be classically stable.

To orient ourselves we reconsider the situation in the {\it basic} model. 
There the curvature decays with an inverse square power like the 
monopole\cite{'tHP} on $\R_3$ and like the latter the asymptotic 
connection behaves as
\[
A_{\mu}^{(\infty)} = \frac{1}{2} g^{-1} \partial_{\mu} g\: ,\qquad g=\hat 
x_{\mu} \sigma_{\mu} ,\: g^{-1}=\hat x_{\mu} \tilde \sigma_{\mu} ,
\]
with $\sigma_{\mu} =(i\vec \sigma , 1)$, and$\tilde \sigma_{\mu} =(-i\vec 
\sigma ,1)$. After passing to the Dirac-string gauge the only term in \r{1} 
which contributes, namely $^{\omega}F_{\mu \nu \rho \sigma}$, decays 
with the {\it eighth} power of $r$ and hence the $4$-dimensional integral 
of it decays with the {\it fourth} power of $r$. It is clear why this decay is 
too strong to result in a Coulmb potential since the latter is characterised 
by its {\it inverse square} behaviour in $4$-dimensions. This leads us 
unambiguously to the criterion that the extension to our {\it basic} model 
must include a term decaying with the {\it sixth} power of $r$, and since 
we expect the asymptotic properties of the {\it extended} model to be the 
same as those of the {\it basic} model \r{12}, i.e. {\it inverse square} 
decaying curvature, the extension must include the 
{\it third} power of the curvature. Inevitably this means that a new 
dimensional constant different from the Higgs vacuum expectation value 
$\eta$ must be introduced. We shall refer to this as criterion (A) below.

The other criterion, (B), which we shall require is that the density in 
question involve no higher powers of the velocity fields than the {\it 
second}. In other words, for fixed $\mu$ and $\nu$, no higher powers of 
$(\partial_{\mu} A_{\nu})^2$ should occur in the new density. This 
criterion is respected by all Skyrme-like models, including the Skyrme 
model\cite{S} as well as the hierarchy of Yang-Mills models\cite{T1} and 
their descendents\cite{T2}. In the context of some more 
phenomenologically motivated considerations, this criterion could be 
relaxed, but this is not the case in the present work. The reason for 
requiring this criterion is that in its absence the definition of the 
canonical momentum-fields would lead to problems in, say the collective 
coordinate quantisation and more specifically in the analysis of the 
fluctuation spectrum when this is eventually performed. While this 
analysis is beyond the scope of the present work, we envisage that it should 
in principle be accessible and hence insist here on this criterion (B).

Having stated our main criteria (A) and (B) for the extension, we introduce 
the other criteria. Obviously this density must be both Lorentz invariant, 
and gauge invariant. In addition it would be an advantage from the 
viewpoint of esablishing the existence of the instanton of the extended 
model, if this density was positive definite by construction. As we shall see 
below, it will not be possible to satisfy this last criterion.

Subject to the constraints of Lorentz and gauge invariances, criterion(A) 
narrows the choices for candidates down to the following three densities:
\[
\mbox {Tr}\: F_{\nu \lambda}F_{\lambda \mu}F_{\mu \nu},\qquad \mbox 
Tr\: F_{\mu \lambda}F_{\lambda \nu}\: ^{*}F_{\mu \nu},\qquad 
\mbox Tr\: (D_{\lambda} F_{\mu \nu})(D_{\lambda} F_{\mu \nu}).
\]
The last of these is equivalent to the non-local effective action density for 
low momentum gluons\cite{DP}\cite{BBZ} $\mbox {Tr} F_{\mu \nu} D^2 
F_{\mu \nu}$, while the other possible term $\mbox {Tr} (D_{\mu} F_{\mu 
\nu})^2$ with the required dimensions is related to the two of the above  
listed terms through
\begin{equation}
\label{34z}
\mbox {Tr} (D_{\mu} F_{\mu \nu})^2 =2\mbox {Tr} F_{\mu \nu}F_{\nu 
\lambda}F_{\lambda \mu} +\frac{1}{2} \mbox {Tr} (D_{\lambda} F_{\mu 
\nu})^2 .
\end{equation}
Invoking the criterion (B) eliminates all but the first of 
these three candidates, namely the extension to our basic model \r{1} must 
be uniquely

\begin{equation}
\label{34}
{\cal L}_3 =\kappa_1^2 \mbox {Tr} F_{\mu \nu} F_{\nu \lambda} 
F_{\lambda 
\mu},
\end{equation}

where $\kappa_1$ is the new constant with the same dimensions as $\eta$. 
This particular dynamical model \r{34} defined on $\R_6$, has appeared 
previously in the literature\cite{Sc}\cite{Y} in a different context where 
self-
dual solutions are found\cite{Sc}\cite{Y} and their stability is 
examined\cite{MaT}. In particular, the fact that criterion (A) is satisfied by 
\r{34} was discussed in some detail in Ref.\cite{Sc}. In the present context 
on $\R_4$, it was first proposed in Ref.\cite{O'BT2}.

Anticipating a result of the next Subsection, namely that the extension of 
\r{1} by the density \r{34} will result in a {\it power} decay for the gauge 
connection function $f(r)$ which in its absence had decayed {\it 
exponentially} 
according to \r{16}, we introduce a second extension to be added to the first 
one \r{34} with the aim of rectifying this situation. This second extension, 
which will introduce yet another dimensional constant, is

\begin{equation}
\label{35}
{\cal L}_2 =\kappa_2^4 \mbox Tr \left( (D_{\mu} \Phi )^2 + \lambda_5 
(\eta^2 +\Phi^2)^2 \right),
\end{equation}

where $\kappa_2$ is this second constant with the dimensions of 
$\eta$, and $\lambda_5$ is an unimportant dimensionless constant which 
will be set equal to $1$ below. The density \r{35} is immediately recognised 
as the usual quadratic Higgs kinetic term plus the usual symmetry 
breaking potential. This term, being entirely $\Phi$ dependent, will have 
no effect on the inter-instanton interactions and is introduced merely to 
restore exponential decay. Clearly it is not possible to add the usual YM term
$F_{\mu \nu}^2$ term to \r{35} as this will render the action 
logarithmically divergent and would invalidate the instanton solution.

We note here that the optional criterion of positive definiteness is {\it not} 
satisfied by \r{34}. This will necessitate some detailed asymptotic and 
numerical analysis of the extended systems carried out below, because the 
lack of positive definiteness of the extension means that the 
topological stability of the instanton of the {\it basic} model \r{1} does not 
automatically guarantee the same for the extended model. Indeed, if the 
coupling constant $\kappa_1$ in \r{34} is allowed to be large, its lack of 
posititvity will destroy the stability of the instanton in the extended model. 
Thus it is imperative that this constant $\kappa_1$ be taken to be small. 
What small means in this context will be determined according to 
quantitative criteria to be applied in Subsection {\bf 3.2} where the 
numerical integrations are described, and the extent to which these 
criteria are met will be discussed in detail in Section {\bf 4}. Here, in 
justification the extension \r{34}, we suffice to note that the value of the 
this density remains positive as long as the asymtotic conditions \r{12} are 
respected as we shall find to be the case in the Subsection {\bf 3.1} 
immediately below.

Summarising, we propose two related extensions of \r{1}. The first is 
obtained by the addition of \r{34} to \r{1}, anmely ${\cal L}_{I}={\cal L}_4 
+{\cal L}_3$, and the second by the addition of \r{34} and \r{35} to \r{1}, 
namely ${\cal L}_{II}={\cal L}_4 +{\cal L}_3 +{\cal L}_2$.

\subsection{Asymptotic solutions}

The one-dimensional subsystems $L_3$ and $L_2$ arising from the 
imposition of spherical 
symmetry on the extensions ${\cal L}_3$ and ${\cal L}_2$, analogous to the 
one-dimensional subsystem \r{9} of \r{1} are, respectively

\begin{equation}
\label{36}
L_3 = \frac{\kappa_1^2}{r} (1-f^2)\left( 3f_r^2 +\frac{1}{r^2} (1-
f^2)^2 \right)
\end{equation}
\begin{equation}
\label{37}
L_2 =\frac{1}{2} \kappa_2^4 r^3 \left( h_r^2 +\frac{3}{r^2} f^2 h^2 
+2\lambda_5 (\eta^2 -h^2)^2 \right).
\end{equation}

It is important to note that the one dimensional density $L_3$ remains 
positive  provided that the asymptotic conditions \r{12} are satisfied by the 
solution, inspite of not being a positive definite quantity by construction. 
We shall find this to be the case below when we solve the Euler-Lagrange 
equations $\delta_f L_I =0$ and $\delta_f L_{II} =0$ in the region $r>>1$. 
This will be the justification of employing \r{34} as an extension.

The Euler-Lagrange equations $\delta_f L_3 =0$ for \r{36}, and, $\delta_f 
L_2 =0$ and $\delta_h L_2 =0$ for \r{37}, are given respectively by

\begin{equation}
\label{38}
\delta_f L_3 =-\frac{6\kappa_1^2}{r} \left( (1-f^2)f_{rr} -f\: f_r^2 
-\frac{1}{r} (1-f^2)f_r +\frac{1}{r^2} (1-f^2)^2 f \right)
\end{equation}
\begin{equation}
\label{38a}
\delta_f L_2 = 3\kappa_2^4 r\: h^2 \: f 
\end{equation}
\begin{equation}
\label{39}
\delta_h L_2 = \kappa^4 r\: \left( 3f^2 h -4\lambda_5 r^2 (\eta^2 -h^2)\: h 
-3rh_r -r^2 h_{rr} \right).
\end{equation}

To start with we dispose of the question of the asymptotic solutions in the 
$r<<1$ region, to both the extended models ${\cal L}_{I}$ and ${\cal 
L}_{II}$. We have verified that the additional terms do not change the 
behaviours \r{18} and \r{19} for the functions $f(r)$ and $h(r)$ 
respectively. There then just remains to find the asymptotic behaviours in 
the $r>>1$ region.

Consider first the extended system ${\cal L}_I ={\cal L}_4 + {\cal L}_3$. The 
$\delta_h L_I =\delta_h L_4 =0$ equations of its one-dimensional subsystem 
are unchanged and are given by \r{11}, while the $\delta_f L_I =\delta_f 
L_4 +\delta_f L_3 =0$ equations are now given by \r{10} and \r{38}. 
Linearising the latter around the asymptotic value of $f(r)=0+F(r)$ we find

\begin{equation}
\label{40}
r^2 \: F_{rr} -r\: F_r+\left( 1-(\frac{\lambda_1 \eta^2}{\kappa_1^2}) 
\right)F =0,
\end{equation}
which has a power {\it decay} solution yielding
\begin{equation}
\label{41}
f(r)\sim r^{1-\sqrt{\lambda_1} \frac{\eta}{\kappa_1}}
\end{equation}
provided that $\sqrt{\lambda_1} \frac{\eta}{\kappa_1} >1$. This is not 
physically unreasonable since the constant $\eta$ is expected to be large, 
while $\kappa_1$ can be taken to be small. In practice we have set $\eta 
=1$ in all our numerical computations and, taken $\lambda_1$ to be of the 
order of unity, while $\kappa_1$ must be taken to be sustantially smaller 
than unity if the lack of positive definiteness of ${\cal L}_3$ \r{34} is not to 
predjudice the instanton solution of the {\it basic} model. Thus, for the 
parameters employed in our numerical computations, the conditions for 
\r{41} to imply a rapid decay are met.

Notwithstanding the fact that we expect a rapid decay \r{41} for the 
function $f(r)$, it would be desirable to extend the model in such a way that 
like the function $h(r)$, the function $f(r)$ also decays exponentially. This 
is because the validity of the approximations that will be made in an 
eventual application of the instanton solutions found here to the 
construction of a dilute instanton gas, rely on the strong localisation of the 
instantons and the best way of meeting this requirement is by exponential 
localisation.

To this end, we consider the extension ${\cal L}_{II} ={\cal L}_4 + {\cal L}_3 
+{\cal L}_2$. In this case, the one-dimensional Euler-Lagrange equations 
$\delta_f L_{II} =\delta_f L_4 +\delta_F L_3 +\delta_f L_2 =0$ and $\delta_h 
L_{II} =\delta_h L_4 +\delta_h L_2 =0$ are given by \r{10}, \r{11}, \r{38} 
and \r{39}. Linearising these around their asymptotic values $f(r)=0+F(r)$ 
and $h(r)=\eta +H(r)$ we find
\begin{equation}
\label{42}
\rho^2 F_{\rho \rho} -\rho F_{\rho} -\rho^4 F=0
\end{equation}
\begin{equation}
\label{43}
\sigma^2 H_{\sigma \sigma} +3\sigma H_{\sigma} -\sigma^2 H=0
\end{equation}
where we have used the dimensionless rescalings $\rho 
=(\frac{\kappa_2^4}{2\kappa_1^2 \eta^2})^{\frac{1}{4}} \eta r$ and 
$\sigma =2\sqrt{2\lambda_5} \eta r$ of the radial variable $r$. Note that 
the notations $\rho$ and $\sigma$ in \r{42} and \r{43} are different from 
those in \r{14} and \r{15}. From the fourth power of $\rho$ in the last term 
of \r{42} it is clear that this equation cannot be brought to the form of a 
modified Bessel equation so we evaluate its asymptotic solution directly, 
yielding $f(r)$ in the $r>>1$ region to be
\begin{equation}
\label{44}
f(r)=\mbox e^{\frac{\kappa_2^2}{2\sqrt{2} \kappa_1 \eta} \eta^2 r^2} ,
\end{equation}
which is clearly exponentially localised. Equation \r{43} on the other hand 
can be brought to the form of a Bessel equation with the solution 
$H(\sigma)=\sigma^{-1}K_1(\sigma)$ yielding

\begin{equation}
\label{45}
h(r)=\eta \left( 1 -\frac{1}{2\sqrt{2\lambda_5} \eta r} K_1 
(2\sqrt{2\lambda_5} \eta r) \right) .
\end{equation}

In summary, we see that in the extended model ${\cal L}_I ={\cal L}_4 + 
{\cal L}_3$, the function $h(r)$ is exponentially localised according to 
\r{17} , while the function $f(r)$ is power localised according to \r{41}. In 
the extended model ${\cal L}_{II} ={\cal L}_4 + {\cal L}_3 +{\cal L}_2$ both 
functions $f(r)$ and $h(r)$ are exponentially localised according to \r{44} 
and \r{45} respectively.

We see that the function $h(r)$ is exponentially localised according to 
\r{17} and \r{45} for both the extended models ${\cal L}_I ={\cal L}_4 + 
{\cal L}_3$ and ${\cal L}_{II} ={\cal L}_4 + {\cal L}_3 +{\cal L}_2$, 
respectively, while the fuction $f(r)$ is power localised according to \r{41} 
for the model ${\cal L}_I ={\cal L}_4 + {\cal L}_3$ and exponentially 
localised according to \r{44} for ${\cal L}_{II} ={\cal L}_4 + {\cal L}_3 
+{\cal L}_2$.

Perhaps the most important result of this Subsection is, that the asymptotic 
behaviours of the function $f(r)$ in both extended models given 
respectively by \r{18},\r{41}, and, \r{18},\r{44}, ensure that the one 
dimensional density \r{36} does not become negative.

\subsection{Numerical solutions}

For the purposes of numerical integrations we have employed the 
following two values of the coupling constant $\kappa_1$, $\kappa1 =0.1$ 
and $0.01$ in\r{34}, which result in appropriately small quantitative 
extensions of the {\it basic} model.

For the extended model ${\cal L}_I ={\cal L}_4 +{\cal L}_3$ given by \r{1} 
and \r{34}, we have integrated equations $\delta_f L_I =\delta_f L_4 + 
\delta_f L_3 =0$ and $\delta_h L_I =\delta_h L_4 =0$ given respectively by 
\r{10}, \r{38}, and by \r{11}, numerically, starting with 
the functions \r{18} and \r{19} near $r=0$ and choosing the numerical 
values of the constants $A$ and $B$ so that $f(r)$ and $h(r)$ 
reach their asymptotic values given by the second members of \r{12} and 
\r{13}.  We have performed this integration only for the dimensionless 
coupling constants $\vec \lambda^{(3)} =(1,1,1,1)$. The 
numerical values of the pair of constants $\{ A, B\}$ are fixed by the 
numerical integrations, and are listed in the first and second colums of 
Table 2. The profiles of the functions $f(r)$ for this extended model are not 
given since they are qualitatively the same as those in Figure 1, while the 
function $h(r)$ is the same as that for the {\it basic} model and is given by 
the $\vec \lambda =(1,1,1,1)$ curve in Figure 2. The numerical values of the 
total actions $S_{I} (\kappa_1 =0.1)=$ and $S_{I} (\kappa_1 =0.1)=$ 
pertaining to these two solutions are listed in the third column of Table 2.

For the extended model ${\cal L}_{II} ={\cal L}_4 +{\cal L}_3 +{\cal L}_2$ 
given by \r{1}, \r{34} and \r{35}, we have employed the same two values of 
the coupling constant $\kappa_1 =0.1,\: 0.01$ as above and for each of these 
the values of the coupling constant $\kappa_2 =0.25$ and $\kappa_2 =0.5$. 
We have integrated the two second order equations $\delta_f L_{II} 
=\delta_f L_4 + \delta_f L_3 +\delta_f L_2 =0$ given by \r{10} \r{38} and 
\r{38a}, and, $\delta_h L_{II} =\delta_h L_4 + \delta_2 L_2 =0$ given by 
\r{11} and \r{39}. Again we have performed this integration only for $\vec 
\lambda =(1,1,1,1)$, and the values of the pair of numerical constants $\{ A, 
B\}$ are fixed by the numerical integration and are listed in the first and 
second columns of Table 3. Again the profiles of the functions $f(r)$ and 
$h(r)$ are not exhibited since they are qualitatively the same as those in 
Figures 1. and 2. The numerical values of the total actions pertaining to 
these two solutions are $S_{II} (\kappa_1 =0.1)=$ and $S_{II} (\kappa_1 
=0.01)=$.

Since the density ${\cal L}_3$ \r{34} is {\bf not} positive definite, the 
stability of the instantons of both the models decsribed by the densities 
${\cal L}I$ and ${\cal L}_{II}$ are not guaranteed by the stability of the 
instanton of the {\it basic model}. We argued above however, that if the 
coupling constant $\kappa_1$ is small enough, the instantons of the 
extended models will nevertheless be stable. To this end, the values of the 
total actions corresponding to each instanton given above are relevant. 
Quantitatively, we seek to demonstrate that the addition of the extension 
densities \r{34} and \r{35} to the {\it basic} density \r{1} does {\bf not} 
result in an appreciable change in the value of the action integrals. The 
detailed comparisons are deferred to Section {\bf 4}.

Before proceding to Section {\bf 4} however, we must perform one further 
numerical operation. From the viewpoint of these quantitative 
comparisons, the difference of the extended model ${\cal L}_{I}$ from the 
{\it basic} model is just the non-positive density ${\cal L}_3$ and hence the 
interesting quantity is the difference of their action integrals. The 
extended model ${\cal L}_{II}$ however is the result of the addition of the 
non-positive density ${\cal L}_3$ to the positive definite density ${\cal 
L}_{III}={\cal L}_4 + {\cal L}_2$ and not to the {\it basic} density alone. The 
interesting quantity in this case therefore is the difference of the action 
integral of ${\cal L}_{II}$ from that of the system described by ${\cal 
L}_{III}$, which we have not studied numerically hitherto. This we now do 
by integrating the Euler-Lagrange equations $\delta_f L_{III} =\delta_f L_4 
+\delta_f L_2 =0$ given by \r{10} and \r{38a}, and $\delta_h L_{III} 
=\delta_h L_4 +\delta_h L_2 =0$ given by \r{11} and \r{39}. We have 
performed the numerical integrations for the value of $\vec \lambda^{(3)} 
=(1,1,1,1)$ only, and for the values of the coupling constant $\kappa_2 
=0.25$ and $\kappa_2 =0.5$. The numerical values of the constants $\{ 
A,B\}$ for these solutions are listed in the first and secon columns of Table 
4, while the values of the total action arelisted in the third column of Table 
4. We do not plot the profiles of the functions $f(r)$ and $h(r)$ in this case 
because they are qualitatively the same as those in Figures 1 and 2.

\section{Summary and discussion}

We have made a detailed numerical study of {\it unit} topological charge 
instanton solutions of the $SO(4)\times U(1)$ Higgs model\cite{O'BT} and its 
extensions\cite{O'BT2}on $\R_4$. The extensions to the {\it basic} model are 
devised such that the dynamics of the extended model would in principle be 
capable of supporting a dilute Coulomb gas of interacting instantons. This 
construction however is deferred to future work, and in the present work 
we have restricted ourselves to establishing the existence of  the classical 
instanton solutions using numerical methods.

In Section {\bf 2}, we have studied the spherically symmetric instanton 
solution of the basic models \r{1} parametrised by a set of dimensionless 
constants $\vec \lambda =(\lambda_1 ,\lambda_2 ,\lambda_3 ,\lambda_4)$, 
for four different such sets $\vec \lambda_i ,i=1,2,3,4$ listed in Table 1. The 
results are exhibited in Table 1 and Figures 1 and 2. The unusual feature of 
the system \r{1} is that for the case $i=3$, namely for $\vec \lambda_3 
=(1,1,1,1)$, where the lower bound on the action is saturated by the 
Bogomol'nyi equations \r{3}-\r{5}, the only solution is the trivial constant 
asymptotic values of the fields. That equations \r{3}-\r{5} are 
overdetermined can clearly be seen by inspection of their spherically 
symmetric restrictions \r{21}-\r{24}, which are four constraints on two 
functions. This feature is shared with the usual Skyrme model\cite{S}, in 
which case however the departure of the soliton solution from the self-dual 
field configuration is fixed by the absolute scale there. Here by contrast 
this departure is depends on the dimensionless coupling constant $\vec 
\lambda_i$, and in this sense our model is more akin to the Abelian Higgs 
model\cite{JbR} on $\R_2$ which even though it {\it does} support self-
dual solutions, the departure from self-duality depends on the value of the 
dimensionless coupling strength $\lambda$ multiplying the Higgs 
potential. $\lambda$ in that case is the analogue of $\vec \lambda_i$ here. 
In both models the value of the action integral increases with an increase 
in the value of these dimensionless coupling strength(s). This is seen to be 
the case from the third column of Table 1, where all the action integrals 
$S(\vec \lambda_i)$ are larger than the topological lower bounds. In 
particular the action integral $S(\vec \lambda_4)$ is larger than the lower 
bound $1$ in that case, while the actions $S(\vec \lambda_i)$ for $i=1,2,3$ 
are each larger than their respective lower bounds $(\lambda_i)_{min} $. 
To demonstrate more quantitatively the departute from the self dual 
configuration, we have plotted the profiles of the one-dimensional 
restriction of the stress-tensor $T_{44}$ for each solution in Figure 4.

In Section {\bf 3} we have given some arguments in favour of our choice 
of extended models and proceded to study these numerically. It turns out 
that our choice for an extension of \r{1} is narrowed down to the cubic 
density \r{34}. This density being non positive definite, there is no 
guarantee that its addition to \r{1} will not invalidate the topological lower 
bound. In that case the extended model would not support an instanton 
solution. We have argued that for small enough coupling of this term, the 
instanton solution persists in the extended model is the dynamics of the 
basic model is robust enough. The numerical data pertaining to this 
extended model ${\cal L}_{I}$ is given in Table 2. We can see that the action 
of ${\cal L}_{I}$ does not differ appreciably from the action of the basic 
model, by comparing the actions $S_{I}(\vec \lambda_3 ,\kappa_1)$ in 
Table 2 with the action $S(\vec \lambda_3)$ in Table 1. Indeed, this 
difference is smaller for the case with the smaller value of the coupling 
constant $\kappa_1$, thus justifying our claim that for small enough 
modifications of the basic model we can expect to have intanton solutions.

The function $h(r)$ parametrising the instanton of the extended model 
${\cal L}_{I}$ is localised exponentially according to \r{17}, but the 
function $f(r)$ is power localised according to \r{41}. This has motivated us 
to consider yet another extension, for which both functions $h(r)$ and 
$f(r)$ are both exponentially localised. This is the model ${\cal L}_{II}$ 
which results from the further addition of the density \r{35} to ${\cal 
L}_{I}$ and whose exponentially decaying solutions are given by \r{44} 
and \r{45}. The numerical data for this model is given in Table 3. To carry 
out the corresponding quantitative checks as in the previous case, we need 
to compare the values of the actions $S_{II}(\vec \lambda_3 , \kappa_1)$ 
with the action integral of the positive definite density defined by the sum 
of \r{1} and the positive definite density ${\cal L}_2$ defined by \r{35}. We 
have denoted this system by ${\cal L}_{III}$ and listed the numerical data 
pertaining to it in Table 4. We must therefore compare the actions 
$S_{II}(\vec \lambda_3 ,\kappa_1)$ in Table 3 with the corresponding 
actions $S_{III}(\vec \lambda_3 ,\kappa_2)$ in Table 4. The qualitative 
conclusions are exactly the same as those for the extended model ${\cal 
L}_{I}$, namely that the action in the extended model ${\cal L}_{II}$ is 
close enough to that of the basic model, that ${\cal L}_{II}$ also supports 
instanton solutions.

Quantitatively, the differences $S_{I}(\vec \lambda_3 ,\kappa_1 
=0.01,\kappa_2 =0) -S(\vec \lambda_3 ,\kappa_1 =0,\kappa_2 =0) =67\times 
10^{-6}$ and $S_{II}(\vec \lambda_3 ,\kappa_1 =0.01, \kappa_2 =0.25) 
-S_{III}(\lambda_3 ,\kappa_1 =0,\kappa_2 =0.25) =68\times 10^{-6}$ are 
practically equal, implying that the system \r{1} is equally robust against 
the addition of \r{34} as against the addition of \r{34} and \r{35}, even 
though in the second case the exponential decay was better satisfied. 
Evidently, this means that the value chosen for the coupling constant 
$\kappa_1$ was sufficiently small so that the extended systems arising 
from the addition of the non-positive definite $F^3$ term \r{34} result in 
new systems that also support instanton solutions. To test the validity of this 
criterion, consider the differences of the corresponding actions for the 
extended models with a larger value of the coupling constant $\kappa_1$. 
Thus $S_{I}(\vec \lambda_3 ,\kappa_1 =0.1,\kappa_2 =0) -S(\vec \lambda_3 
,\kappa_1 =0,\kappa_2 =0) =6665\times 10^{-6}$ and $S_{II}(\vec \lambda_3 
,\kappa_1 =0.1, \kappa_2 =0.25) -S_{III}(\lambda_3 ,\kappa_1 =0,\kappa_2 
=0.25) =6685\times 10^{-6}$, which are again nearly equal to each other but 
are very considerably larger than the former pair of differences. This 
completes our justification of the extended models ${\cal L}_{I} ={\cal L}_4 
+{\cal L}_3$ and ${\cal L}_{II} ={\cal L}_4 +{\cal L}_3 +{\cal L}_2$.

Having established that the {\it extended} models also can support 
instantons with the same asymptotic properties as the instantons of the {\it 
basic} model, the remaining task is to compute the volume integral

\begin{equation}
\label{46}
S_{int} = \int d^4 x \: \; \sum_{a,b,c} \mbox {Tr} F_{\mu \nu}(x-x_a) F_{\nu 
\lambda}(x-x-x_b) F_{\lambda \mu}(x-x_c)
\end{equation}

in the Dirac string gauge. Clearly $S_{int}$ in \r{46} is the {\bf only} 
nonvanishing contribution to the action due to inter-instanton 
interactions since in the Dirac string gauge the asymptotic contributions 
of all Higgs dependent terms vanish and the density of the only other Higgs 
independent term in either extended action density coming from $F_{\mu 
\nu \rho \sigma}$ also vanishes according to our arguments at the end of 
Subsection {\bf 3.2}, namely \r{33}. The integrand in \r{46}, which is 
readily calculated by substituting \r{27} in \r{46}, is a quantity of 
considerable complexity and since we will not perfom the integration here 
we do not write it down. If one set any two of the summation indices $a,b,c$, 
in \r{46} the same, namely restricted to "two neighbour" interactions, then 
the integrand would become singular at the origin in the usual way. Thus 
the interaction term will have to take account of "three neighbour" 
interactions. This situation may be changed by replacing the cubic 
extension term \r{34} in \r{46} by the alternative density $F_{\mu \nu} 
D^2 F_{\mu \nu}$ appearing in \r{34z}

\begin{equation}
\label{47}
S_{int} = \int d^4 x \: \; \sum_{a,b} \mbox {Tr} D_{\lambda} F_{\mu \nu}(x-
x_a) D_{\lambda} F_{\mu \nu}(x-x-x_b),
\end{equation}

which is a "two neighbour" interaction. The details of these contributions 
to the action as well as their consequences in the resulting effective theory 
of Strong interactions will be investigated and given elsewhere.

{\bf Acknowledgements}We thank J. Burzlaff for his cooperation at the 
early stages of this work.

\newpage

TABLES:

Table 1

\begin{tabular}{|l|c|c|c|} \hline \hline $ \lambda_i $ & $ A $ & $ B $ & $ 
S(\vec \lambda_i) $ \\ \hline 0.5 & 0.300044 & 0.653908 & 1.023269 \\
0.8 & 0.324460 & 0.650557 & 1.611897 \\
1.0 & 0.335691 & 0.648918 & 2.002486 \\
1.2 & 0.344599 & 0.647561 & 2.392250 \\ \hline \hline 

\end{tabular}

Table 2

\begin{tabular}{|l|l|l|l|} \hline \hline \multicolumn{4}{|c|}{ $\kappa_2 = 
0$} \\ \hline $ \kappa_1 $ & $ A $ & $ B $ & $S_I (\vec \lambda_3 
,\kappa_1)$ \\ \hline 0.01 & 0.335669 & 0.648919 & 2.002553\\
0.1 & 0.333488 & 0.648989 & 2.009151\\ \hline 

\end{tabular}

Table 3

\begin{tabular}{|l|l|l|l|} \hline \hline \multicolumn{4}{|c|}{ $\kappa_2 = 
0.25 $} \\ \hline $ \kappa_1 $ & $ A $ & $ B $ & $ S_{II}(\vec \lambda_3 
,\kappa_1) $ \\ \hline 0.01 & 0.336761 & 0.651287 & 2.010352 \\ 0.1 & 
0.334577 & 0.651353 & 2.016969 \\ \hline \multicolumn{4}{|c|}{ $\kappa_1 = 
0.50 $} \\ \hline $ \nu_i $ & $ A $ & $ B $ & $ S_{II}(\vec \lambda_3 
,\kappa_1) $ \\ \hline 0.01 & 0.350511 & 0.679313 & 2.108078\\
0.1 & 0.348347 & 0.679341 & 2.115000\\ \hline 

\end{tabular}

\bigskip

Table 4

\begin{tabular}{|l|l|l|l|} \hline \hline \multicolumn{4}{|c|}{ $\kappa_1 = 0 
$} \\ \hline $ \kappa_2 $ & $ A $ & $ B $	& $ S_{III}(\vec \lambda_3 
,\kappa_2) $ \\ \hline
$0.25$ & 0.336783 & 0.651286 & 2.010284\\ $0.50$ & 0.350533 & 0.679313 & 
2.108006\\ \hline \end{tabular}

\bigskip
\bigskip

\noindent
FIGURE CAPTIONS:

\noindent
Figure 1. Profiles of the function $f(r)$ for the systems characterised by 
$\vec \lambda_1 =(0.5,0.5,0.5,0.5)$, $\vec \lambda_2 =(0.8,0.8,0.8,0.8)$, $\vec 
\lambda_3 =(1,1,1,1)$, $\vec \lambda_3 =(1.2,1.2,1.2,1.2)$ respectively from 
right to left.

\noindent
Figure 2. Profiles of the function $h(r)$ for the systems characterised by 
$\vec \lambda_1 =(0.5,0.5,0.5,0.5)$, $\vec \lambda_2 =(0.8,0.8,0.8,0.8)$, $\vec 
\lambda_3 =(1,1,1,1)$, $\vec \lambda_3 =(1.2,1.2,1.2,1.2)$. All curves 
strongly overlapping.

\noindent
Figure 3. Profiles of the action densities of the solutions to the systems 
characterised by $\vec \lambda_1 =(0.5,0.5,0.5,0.5)$, $\vec \lambda_2 
=(0.8,0.8,0.8,0.8)$, $\vec \lambda_3 =(1,1,1,1)$, $\vec \lambda_3 
=(1.2,1.2,1.2,1.2)$ respectively in order of the heights of the curves.

\noindent
Figure 4. Profiles of $T_{44}$ given by \r{20} for the systems characterised 
by $\vec \lambda_1 =(0.5,0.5,0.5,0.5)$, $\vec \lambda_2 =(0.8,0.8,0.8,0.8)$, 
$\vec \lambda_3 =(1,1,1,1)$, $\vec \lambda_3 =(1.2,1.2,1.2,1.2)$ respectively 
in order of the lowest to highest curves.


\newpage


\begin{thebibliography}{99}

\bibitem{BPST} A.A. Belavin, A.M. Polyakov, A.S. Schwarz and Yu.S. 
Tyupkin, Phys. Lett. {\bf B59} (1975) 85.

\bibitem{P} A.M. Polyakov, Nucl. Phys. {\bf 120} (1977) 249.

\bibitem{'tHP} G. 'tHooft, Nucl. Phys.{\bf 79} (1974) 276; A.M. Polyakov, JETP 
Lett. {\bf 20} (1974) 194.

\bibitem{B-KT} V.L. Berezinsky, JETP {\bf 30} (1970) 493; J.M. Kosterlitz and 
D.J. Thouless, J. Phys. {\bf C6} (1973) L97.

\bibitem{O'BT}G.M. O'Brien and D.H. Tchrakian, Mod. Phys. Lett. {\bf A4} 
(1989) 1389.

\bibitem{T1} D.H. Tchrakian, "Yang-Mills Hierarchy", Int. J. Mod. Phys. A 
(Proc. Suppl.) {\bf 3A} (1993) 584, and references therein.

\bibitem{T2} D.H. Tchrakian, "Skyrme-like Models in Gauge Theory", in 
"Constraint Theory and Quantisation Methods", eds. F. Colomo, L. Lusanna 
and G. Marmo, World Scientific, Singapore, 1994, and references therein.

\bibitem{ST} T.N. Sherry and D.H. Tchrakian, Phys. Lett. {\bf B 295} (1992) 
237.

\bibitem{JbR} L. Jacobs and C. Rebbi, Phys. Rev. {\bf B19} (1979) 4486.
    
\bibitem{S} T.H.R. Skyrme, Nucl. Phys. {\bf 31} (1962) 556.

\bibitem{TC} D.H. Tchrakian and A. Chakrabarti, J. Math. Phys. {\bf 32} 
(1991) 2532.

\bibitem{JR}R. Jackiw and C. Rebbi, Phys. Rev. Lett. {\bf 37} (1976) 172; C.G. 
Callan, R.F. Dashen and D.J. Gross, Phys. Lett. {\bf B63} 334.

\bibitem{P2} see A.M. Polyakov, "Gauge Fields and Strings", Harwood 
Academic Publishers, Chur, 1987.

\bibitem{MT} Zh-Q. Ma and D.H. Tchrakian, Lett. math. Phys. {\bf 26} (1992) 
111.

\bibitem{WY} T.T. Wu and C.N. Yang, J. Math. Phys. {\bf 15} (1974) 53.

\bibitem{Sc}C. Saclioglu, Nucl. Phys. {\bf B227} (1986) 487.

\bibitem{Y} K. Fujii, Lett. math. Phys. {\bf 12} (1986) 363.

\bibitem{MaT} Zh-Q. Ma and D.H. Tchrakian, Lett. Math. Phys. {\bf 19} 237.

\bibitem{O'BT2} G.M. O'Brien and D.H. Tchrakian, "Exponentially localised 
instantons in a hierarchy of Higgs models", in "Geometry of Constrained 
Dynamical Systems", ed. J.M. Charap, Cambridge University Press, 
Cambridge, 1995.

\bibitem{DP} D.I. Diakonov and V. Yu. Petrov, Phys. Lett. {\bf B242} (1990) 
425.

\bibitem{BBZ} M. Baker, J.S. Ball and F. Zachariasen, Nucl. Phys. {\bf B229} 
(1983) 445.


\end{thebibliography}
\end{document}